# Coded aperture correlation holographic microscope for single-shot quantitative phase imaging with extended field of view

NATHANIEL HAI[*] AND JOSEPH ROSEN

*School of Electrical and Computer Engineering, Ben-Gurion University of the Negev. P.O Box 653, Beer-Sheva 8410501, Israel*
*\*mirilasn@post.bgu.ac.il*

**Abstract:** Recently, a method of recording holograms of coherently illuminated three-dimensional scene without two-wave interference was demonstrated. The method is an extension of the coded aperture correlation holography from incoherent to coherent illumination. Although this method is practical for some tasks, it is not capable of imaging phase objects, a capability that is an important benefit of coherent holography. The present work addresses this limitation by using the same type of coded phase masks in a modified Mach-Zehnder interferometer. We show that by several comparative parameters, the coded aperture-based phase imaging is superior to the equivalent open aperture-based method. As an additional merit of the coded aperture approach, a framework for increasing the system's field of view is formulated and demonstrated for both amplitude and phase objects. The combination of high sensitivity quantitative phase microscope with increased field of view in a single camera shot holographic apparatus, has immense potential to serve as the preferred tool for examination of biological tissues and micro-organisms.

## 1. Introduction

Quantitative phase imaging is a label-free method for capturing the complex wavefront originating from transparent samples, and converting it into their corresponding spatial optical thickness (OT) map [1]. This powerful tool enables three-dimensional (3D) non-destructive imaging, in a nanoscale sensitivity, and enables reconstructing the morphology of phase objects that are otherwise invisible to the naked eye. Quantitative phase imaging finds a wide range of applications including stain-free biological cell imaging [2], non-destructive quality tests of optical elements [3] and surface profilometry [4].

In the x-ray regime, indirect imaging using coded aperture masks is a well-known approach to increase the signal-to-noise ratio (SNR), field of view (FOV) and the power efficiency associated with conventional imaging systems [5,6]. Coded aperture imaging involves the recording of a chaotic pattern formed on the sensor plane by a superposition of many scattered signals, each of which originated from a different region of the modulating mask. The recorded pattern must be further processed digitally in order to reconstruct the final image in a deconvolution process. In the optical domain, an efficient, rapid and well-established method to reconstruct the desired image is by cross-correlating the recorded signal with a characteristic pattern of the system in a framework termed coded aperture correlation holography (COACH) [7]. COACH based systems possess the same imaging qualities as an equivalent conventional direct imaging system, in addition to the capability of imaging 3D scenes by three [8], two [9], or even by a single camera shot [10,11]. Originally, COACH was invented as an incoherent, self-interference digital holography technique and was tremendously simplified to coaxial optical setups. Apparently, in coded aperture systems, the information regarding the object's location is encoded also in the magnitude of the signal and not only in its phase. Therefore, a later improvement in COACH was the implementation of holographic apparatus that records incoherent holograms without two-beam interference [8]. The interferenceless feature of COACH has motivated several additional improvements in terms of the imaging capabilities and image qualities. Among the various improvements, there are extended FOV [12], enhanced spatial resolution [13,14], imaging through a scattering layer [15,16] and imaging through partial aperture [17,18].

Recently, we implemented the interferenceless COACH in the coherent optical regime. Consequently, a coherently illuminated 3D scene can be reconstructed from a hologram recorded without interference between object and reference beams [19]. In this particular implementation, we had to accommodate the incoherent nature of COACH systems to the coherent light source. This adaptation was achieved by a unique design of the coded phase mask (CPM) which modulates the object signal. By designing the impulse response of the system as randomly and sparsely distributed light dots, we not only could treat the given coherent system as an incoherent and analyze it as such, but we also earn inherent optimization parameters to enhance the quality of reconstructed images. An advanced coherent COACH with the ability to multiplex more than one hologram per one camera shot has been proposed recently in [20].

In the present study, we aim to improve the coherent COACH system capabilities toward the mission of quantitative phase imaging. The proposed phase imager maintains an important feature in imaging of dynamic phase samples; the acquisition of each hologram is done by a single camera shot. Nevertheless, we show that the COACH approach can be more accurate, and more immune from noise, than an equivalent conventional method. Moreover, the proposed COACH is used to increase the FOV of an equivalent regular system by a factor of three without the need of a special calibration procedure or a scanning operation. The integration of a quantitative phase microscope with the significant advantages of a coded aperture optical system makes the suggested coherent sparse COACH (CS-COACH) a serious candidate for non-invasive and label-free inspection of low power signals emitted from biological samples.

## 2. Methodology

### 2.1. Coherently illuminated COACH

COACH is a two-step process for imaging an incoherently illuminated 3D scene from a single viewpoint. First, the optical system records a two-dimensional (2D) digital hologram into the computer. Then, each 2D transverse plane from the observed 3D scene can be digitally reconstructed from the recorded 2D hologram. Formally, 2D digital convolution-based hologram $H_{OBJ}(\bar{r})$ of an object $I_{OBJ}(\bar{r},z)$ is given by,

$$H_{OBJ}(\bar{r}) = \int I_{OBJ}(\bar{r},z) * t(\bar{r},z) dz, \qquad (1)$$

where $*$ is a 2D convolution, $(\bar{r},z)=(x,y,z)$ are the system coordinates and $t(\bar{r},z)$ is the point spread hologram (PSH) of the optical system which was synthesized to be bi-polar real [9,10], or general complex function [8]. In the second stage of the process, each $z_j$ plane of the 3D image $I_{IMG}(\bar{r},z)$ is reconstructed by digital 2D cross-correlation of the object hologram with a PSH-related reconstructing function $R(\bar{r},z)$ as follows,

$$I_{IMG}(\bar{r},z_j) = H_{OBJ}(\bar{r}) \otimes R(\bar{r},z_j), \qquad (2)$$

where $\otimes$ is 2D correlation.

Recently, our group found a certain type of PSHs with the superiority in terms of the image quality and of the apparatus robustness [19-21]. These PSHs are made of a group of isolated light dots distributed chaotically over the entire image plane of the system. Consequently, the object hologram $H_{OBJ}(\bar{r})$ is ensemble of image replications of the object distributed in accordance of the PSH distribution.

Lately, we extended COACH to the regime of coherent illumination by enforcing a simple condition [19]. Since the reconstruction is carried out by a cross-correlation process, the necessary condition on the random dots, unique to coherent illumination, is that the distance between any two adjacent points should be larger than the image of the input object. In this manner, there is no overlapping and no interference between the different image replications. The number of light dots in the PSH is determined experimentally, subject to the non-overlapping condition, to maximize both the PSH complexity and its power efficiency at the same time. This ensures optimal imaging performance in terms of SNR and visibility, simultaneously [19]. The intensity at the image plane is given by,

$$\left| A_{In}(\bar{r}) * H_{PSH} \right|^2 = \left| A_{In}(\bar{r}) * \sum_k a_k \delta(\bar{r} - \bar{r}_k) \right|^2 = \left| \sum_k a_k A_{In}(\bar{r} - \bar{r}_k) \right|^2$$
$$= \sum_k |a_k|^2 |A_{In}(\bar{r} - \bar{r}_k)|^2 = |A_{In}(\bar{r})|^2 * \sum_k |a_k|^2 \delta(\bar{r} - \bar{r}_k), \quad (3)$$

where $H_{PSH}$ is the PSH of random dots, $A_{In}(\bar{r})$ is the complex amplitude of the system input satisfying the relation $I_{OBJ}(\bar{r}) = |A_{In}(\bar{r})|^2$, $\delta(\cdot)$ is the delta function, $k$ is the index of the dots and $a_k$'s are complex valued constants. The design of an impulse system response as sparse chaotic dots is possible due to a CPM utilized as the system aperture, and enables the interferenceless coherent imaging system [19]. Although this system is efficient in some cases, it has two major limitations arising from the optical design. First, the demand for non-overlap between the object replications at the image plane limits the system's FOV and second, the interferenceless feature of the system implies that the phase information of the object cannot be detected. In the present proposed CS-COACH system we address these two limitations.

### 2.2. Extending the FOV

In the following we assume that the FOV of a given optical system is limited by the size of the camera's active area. For an active area size of $B \times B$ the system FOV equals to $S \times S$, where $S = B/M_T$ and $M_T$ is the system's transverse magnification. In order to increase the effective FOV of the system, we must overcome the image sensor limitation. By displaying the designated CPM on a spatial light modulator (SLM), we generate a PSH of randomly distributed dots at the image plane, some of them are outside the active area of the image sensor. In this way, the information regarding objects outside the FOV of the regular system can be captured by the image sensor. The limitation of this FOV extension technique is due to the finite extent of the PSH, which arises from the maximal scattering angle of the SLM. In this study, we have implemented a threefold increase of the FOV in both transverse axes, in comparison to a regular system.

Given that the active area of the image sensor is $B \times B$, and of the CPM is $3B \times 3B$, the CPM generates a PSH consisting 9 sub-PSHs, each of which having different array of randomly distributed dots. The area ratio between the sensor and the CPM is determined for the algorithm of generating the CPM and can be changed in the operating optical setup without any change in the system FOV. This last statement is valid as long as the CPM is displayed in the Fourier plane of a telescopic system of two lenses. In such system, the focal length of the rear lens determines the area ratio between the sensor and the CPM, and the ratio between the focal lengths of the two lenses determines the area ratio between the sensor and the input plane. The overall PSH is given by,

$$H_{PSH} = \sum_{m=-1}^{1}\sum_{n=-1}^{1} \tilde{H}_{PSH,m,n}(\bar{r}-\bar{B}_{m,n}) = \sum_{m=-1}^{1}\sum_{n=-1}^{1}\sum_{k}^{K_{m,n}} a_{m,n,k}\delta(\bar{r}-\bar{r}_{m,n,k}-\bar{B}_{m,n}), \qquad (4)$$

where $K_{m,n}$ is the number of dots at the $(m,n)$ sub-PSH and $\bar{B}_{m,n}=(mB,nB)$ is the displacement vector of the different sub-PSHs from the origin of the image plane. It is important to note that each of those sub-PSHs is related to a different region of the threefold increased FOV of the system. For a signal outside the FOV of the regular system, the related sub-PSH on the image plane is translated accordingly to be captured by the image sensor. For example, suppose we have placed an object $I_{OBJ}(\bar{r}-\bar{B}_{1,-1}/M_T)$ at the object plane. Although this object is located outside the FOV of the regular system, it induces randomly distributed replications of the image on the image sensor of the size $B\times B$, such that the recoded signal is,

$$I_{CCD}(\bar{r}) = I_{OBJ}(\bar{r}) * \sum_{k}^{K_{1,-1}} a_{1,-1,k}\delta(\bar{r}-\bar{r}_{1,-1,k}). \qquad (5)$$

The $3B\times 3B$ CPM is computed using a modified Gerchberg-Saxton algorithm (GSA) [22] illustrated in Fig. 1. The use of GSA is possible herein due to the Fourier relation between the CPM and the camera plane achieved by locating the CPM in the front, and the camera in the rear, focal planes of a spherical lens ($L_F$ in Fig. 2). An initial random phase propagates from the CPM plane to its Fourier related plane, where the desired intensity of randomly distributed dots over the camera plane is enforced. The intensity on the camera plane is constrained to be the chaotic distributed dots in the entire nine sub-PSHs, whereas the phase is used as the degree of freedom. The constraint on the CPM plane is the group of phase-only matrices according to the property of the used phase-only SLM. When the final CPM is displayed on the SLM, the square magnitude of the ensemble of dots becomes the PSH, such that object located inside or outside the regular system FOV independently creates an ensemble of image replications over the active area of the image sensor.

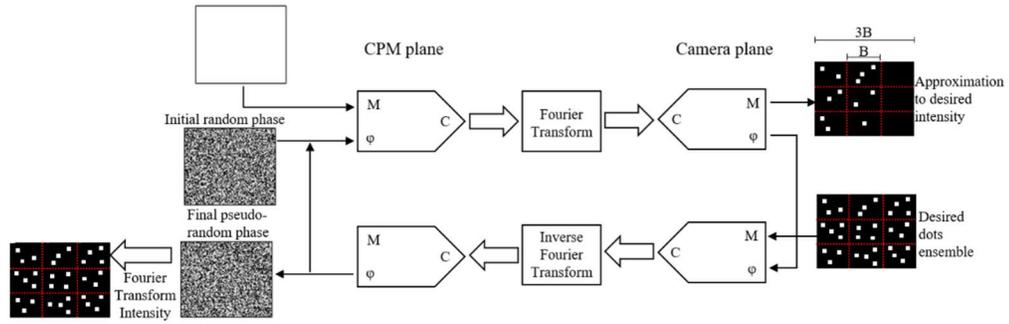

Fig. 1. Flow chart of the modified GSA used for rendering the pseudorandom CPM that generates the extended ensemble of dot pattern on the image plane.

### 2.3. Phase extraction from off-axis hologram

One of the practical ways to image the phase of a given object quantitatively is by encoding the phase to a hologram using an interference process. In COACH, this interference must be executed between the object wave, containing image replications of the input object, and a reference plane wave, that does not carry any information regarding the object. The resulting interferogram is,

$$\left| A_{In}(\bar{r}) * \sum_k a_k \delta(\bar{r} - \bar{r}_k) + A_r \exp(i 2\pi |\bar{r}| \sin\theta / \lambda) \right|^2$$

$$= \left| A_{In}(\bar{r}) * \sum_k a_k \delta(\bar{r} - \bar{r}_k) \right|^2 + |A_r|^2 + 2\sum_k |A_{In}(\bar{r})||a_k||A_r| \cos\left[ 2\pi |\bar{r}| \sin\theta / \lambda - \varphi(\bar{r}) - \beta_k + \alpha \right] * \delta(\bar{r} - \bar{r}_k),$$

(6)

where $A_r$ is the complex amplitude of the reference beam, $\alpha$ is the phase of $A_r$, $\varphi(\bar{r})$ is the phase of the object, $\beta_k$ is the phase of the $k$-th PSH dot, and $\theta$ is the angle between the two interfering waves. $\varphi(\bar{r})$ is related to the OT of the object via the equation OT = $\varphi\lambda/2\pi$, where $\lambda$ is the illumination wavelength in the free space. In order to extract the desired object phase, we use the angle $\theta$ which induces a separation between the spectral terms in the Fourier domain. Consequently, the terms $A_{In}^*(\bar{r}) A_r * \sum_k a_k^* \delta(\bar{r} - \bar{r}_k)$ and $A_{In}(\bar{r}) A_r^* * \sum_k a_k \delta(\bar{r} - \bar{r}_k)$ are located on different sides of the spatial-frequency domain such that each one of them contains the complex wavefront of the object. The above-mentioned process of filtering out the unwanted signal from the interferogram assumes that the optical setup is well designed and properly aligned [23]. Moreover, it is assumed that the $\theta$ angle between the object and reference beams induces enough separation between the central zero-order terms and the interference terms [24]. After the bias terms and the twin image of Eq. (6) are eliminated, the filtered signal contains image replications of the input object wavefront. This complex amplitude can be cross-correlated with a phase-only filtered version of the system's PSH, $H'_{PSH}$, to reconstruct the object image according to Eq. (2) as follows,

$$I_{IMG} = \left[ A_{In}(\bar{r}) A_r^* * \sum_k a_k \delta(\bar{r} - \bar{r}_k) \right] \otimes H'_{PSH}$$

$$= \mathfrak{F}^{-1} \left\{ \mathfrak{F}\left[ A_{In}(\bar{r}) A_r^* * \sum_k a_k \delta(\bar{r} - \bar{r}_k) \right] \exp[-j\Phi(\bar{\rho})] \right\} \quad (7)$$

$$= \mathfrak{F}^{-1} \left\{ \mathfrak{F}\left[ A_{In}(\bar{r}) A_r^* \right] |h(\bar{\rho})| \exp[j\Phi(\bar{\rho})] \exp[-j\Phi(\bar{\rho})] \right\}$$

$$= A_{In}(\bar{r}) A_r^* * \mathfrak{F}^{-1}\{|h|\} \propto A_{In}(\bar{r}) = |A_{In}(\bar{r})| \exp[j\varphi(\bar{r})],$$

where $\bar{\rho} = (\nu_x, \nu_y)$ is the location vector in the spatial frequency domain, $\mathfrak{F}\{\}$ and $\mathfrak{F}^{-1}\{\}$ denotes a 2D Fourier and its inverse transforms, respectively. The Fourier transform of the PSH is given by,

$$\mathfrak{F}\left\{ \sum_k a_k \delta(\bar{r} - \bar{r}_k) \right\} = |h(\bar{\rho})| \exp[j\Phi(\bar{\rho})]. \quad (8)$$

The equivalent sign in the last line of Eq. (7) is valid under the assumptions that $\mathfrak{F}\{|h|\}$ is a delta-like function, which is correct as long as the dots distribution of the PSH is random, and the reference wave is approximately a plane wave. The extracted phase from the complex wavefront reconstruction of Eq. (7) is given to $2\pi$ ambiguities due to the cyclic nature of the trigonometric functions and should be solved by a phase unwrapping algorithm in cases where the object's OT is larger than the illumination wavelength [25].

### 2.4. Experimental setup

To demonstrate the CS-COACH capabilities, we employed a modified Mach-Zehnder interferometer as shown in Fig. 2. A HeNe laser beam (Melles-Griot, Max. output power 75mW @ $\lambda$ = 632.8nm) is split by a beam-splitter (BS$_1$) to reference arm and object arm. In the object arm, the spatial information from the object plane is imaged to infinity by a microscope objective, (Newport M-40X NA = 0.65) and relayed towards the SLM by an optical relay consists of $L_1$ ($f_1$ = 60mm, $D$ = 50.8mm) and $L_2$ ($f_2$ = 200mm, $D$ = 50.8mm). The object spatial spectrum is modulated by the CPM displayed on the SLM (Holoeye PLUTO, 1920×1080 pixels, 8μm pixel pitch, phase-only modulation) and is then Fourier transformed back into the spatial domain by the Fourier lens $L_F$ ($f_F$ = 200mm, $D$ = 50.8mm). The beam from $L_F$ is interfered on the image sensor with the reference beam coming via $BS_3$. In the reference arm, the beam is expanded and spatially filtered using two lenses $L_3$ ($f_3$ = 50mm, $D$ = 25.4mm) and $L_4$ ($f_4$ = 400mm, $D$ = 50.8mm) and a pinhole (Thorlabs, 2μm diameter) in a Keplerian beam expander formation, to achieve full overlap with the image replications from the object arm. Interference pattern between the signals of the object and reference arms is recorded by the image sensor (Thorlabs 8051-M-USB, 3296 × 2472 pixels, 5.5μm pixel pitch, monochrome) with a slight relative angle between the two signals, induced by a small angle tilt of $BS_3$.

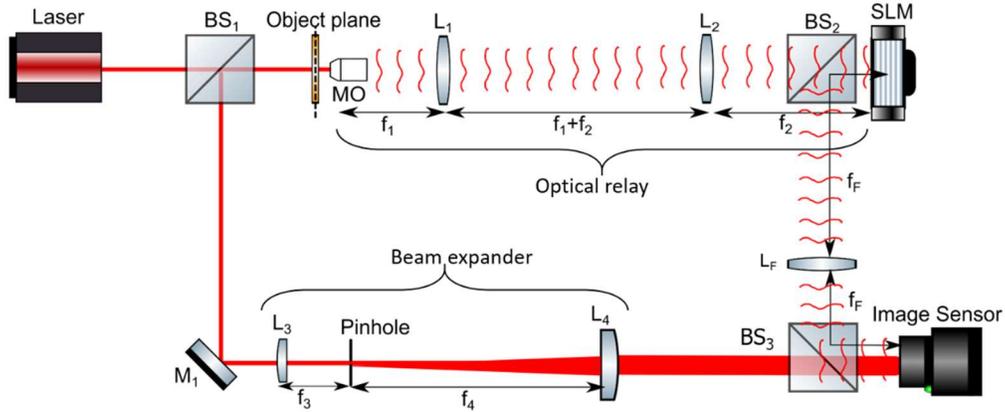

Fig. 2. Optical configuration of the CS-COACH. MO: microscope objective, M: mirror, BS: beam-splitter, SLM: spatial light modulator.

## 3. Results

### 3.1 Extended FOV

Previous attempts to extend the FOV of COACH systems have suffered from substantial background noise in the reconstructed images, forcing the experimenter to decrease the temporal resolution by averaging over several independent reconstructions [12]. To avoid this noise, in the current technique, we use the mentioned-above version of COACH in which the PSH is composed from randomly distributed dots and hence better imaging performance in terms of SNR can be achieved. Combining this approach with off-axis holography enables to acquire image of extended FOV by a single camera shot.

Fig. 3 illustrates the imaging process with and without the FOV extension. Fig. 3(a) depicts the interference pattern between the signals from the object and reference arms. Three input objects were placed at the object plane – two of them outside the regular system FOV (digits 1 and 4) and a single object within the regular system FOV (digit 7). As expected, the emerging interferogram consists of randomly distributed replications of all the three objects along with diagonal fringes corresponding to the relative angle between the two beams of the interferometer. This relative angle induces a separation between the various terms in the spectrum domain of the interferogram. Thus, the unwanted information can be eliminated such that only the desired replications of the input objects are remaining, as illustrated in Fig. 3(b). Object hologram, containing threefold improved FOV in both lateral axes, was generated by zero-padding the filtered interferogram to match the size of the PSH. It should be noted that the PSH is generated digitally in the computer and is not recorded optically as a point response as done in the past [19]. In other words, the correct distribution of the dots is determined during the CPM generation by the GSA, and only the size of the synthetic PSH should be adjusted to match the active area size of the image sensor. The synthetic PSH is a significant advantage of the CS-COACH system over the other COACH methods because a process of optical calibration can be avoided. After tailoring the synthetic PSH, its size becomes three times bigger than the active matrix of the imaging sensor in each axis. The extended FOV image of the original scene can be readily reconstructed by cross-correlating the object hologram [zero padded version of Fig. 3(b)] with the synthetic PSH as shown in Fig 3(c). It can be seen that the two objects that were placed outside the regular system FOV are reconstructed along with the object that was placed within the regular system FOV. The impact of using the coded aperture for increasing the system's FOV is further demonstrated in Fig. 3(d), which describes the result of cross-correlating the original size object hologram [Fig. 3(b)] with the regular PSH containing only the sub-PSH corresponding to the FOV of a regular system. This sub-PSH reconstructs only the information from the regular system FOV, whereas the information from the area outside that FOV is completely lost.

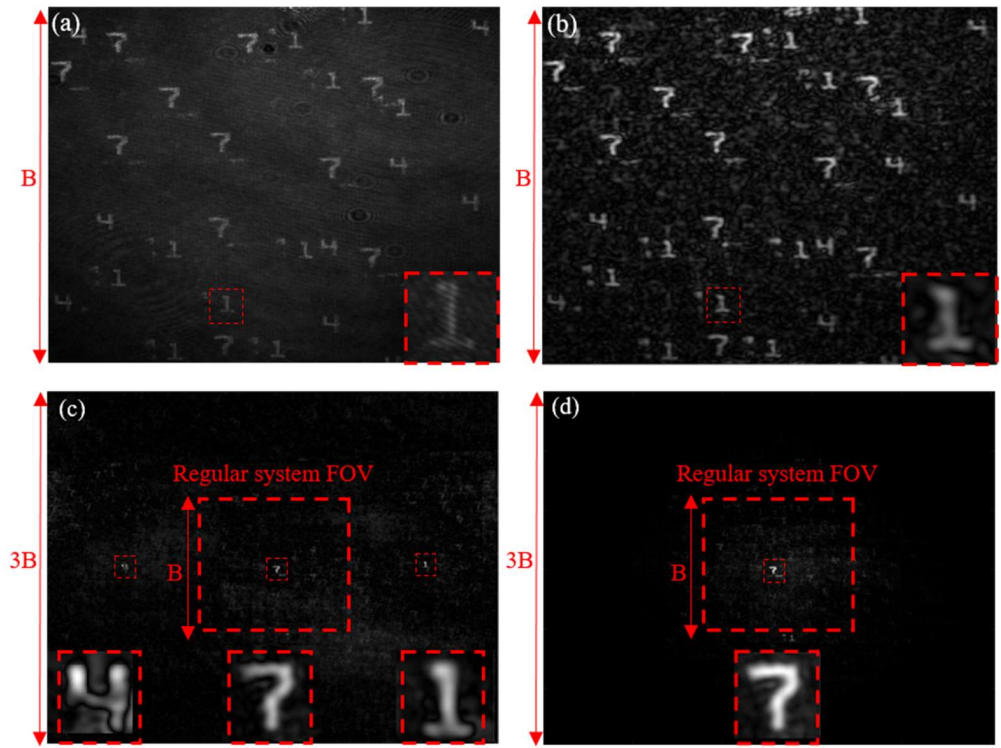

Fig. 3. Extended FOV in CS-COACH: (a) interference pattern between the object arm signal and the reference arm signal as recorded by the image sensor, and its (b) filtered hologram free of bias terms and twin image. (c) Reconstruction of the input scene image, containing objects outside the regular system FOV. This image is obtained by cross-correlation between the object hologram in (b) and the synthetic PSH. (d) Reconstruction of the input scene image, obtained by cross-correlation between the object hologram in (b) and a regular PSH of the size of the camera active area. Consequently, the system FOV is limited to $B/M_T$.

*3.2 Quantitative phase imaging*

In the following, we demonstrate the performances of the CS-COACH as a quantitative phase imaging method. To this end, we placed a 1951 USAF phase resolution chart (Quantitative Phase Microscopy Target, Benchmark Technologies) at the front focal plane of the microscope objective. The refractive index of the sample is 1.52 and the features height upon the substrate is 250 nm, which translates to phase difference of 1.29 radians between the target elements and their substrate. Therefore, if the object's phase map is accurately obtained, phase unwrapping procedure is not necessary. Extraction of the object's phase map was carried out twice, first by displaying an appropriate CPM on the SLM, hence using the proposed CS-COACH method, and second by displaying a constant phase mask on the SLM, thus using a standard Mach-Zehnder interferometer in open aperture off-axis holography. Fig. 4 describes the phase reconstruction procedure using CS-COACH and the comparison between the results of the two approaches. Fig. 4(a) is the recorded off-axis hologram using CS-COACH in which 13 image replications can be observed. One can notice from the inset of Fig. 4(a) that the target elements and their background are transparent to light although the phase of the hologram fringes are different in the two areas. Fig. 4(b) depicts the phase of the filtered object hologram after subtraction of the phase of an equivalent sample-free filtered hologram. Background phase subtraction is needed in order to calibrate the system to the phase noise originating from uncontrolled aberrations and misalignments [for example $\alpha$ and $\beta_k$'s from Eq. (6)]. The result of Fig. 4(b) is replications of the quantitative phase variation of the observed object. In order to reconstruct the entire phase information of the original object, the filtered object hologram is cross-correlated with a phase-only filtered version of the PSH, which was generated in the computer. The phase of the reconstructed image using CS-COACH is illustrated in Fig. 4(c). Fig. 4(d) is the extracted phase of the object after a similar filtering procedure, without the cross-correlation process, of the open aperture off-axis hologram acquired using the Mach-Zehnder interferometer and when a constant phase is displayed on the SLM (recorded off-axis hologram by this Mach-Zehnder interferometer is not shown). Qualitative comparison between the reconstructed phase maps of the two methods shows that the phase reconstruction using the CS-COACH method exhibits sharper edges of the phase elements, indicating that CS-COACH result is closer to the reality of sharp steps in the surface height. Moreover, quantifying the mean square errors (MSEs) of the square element phase of the resolution target (blue dashed square) and the substrate phase (yellow dashed rectangular) of Figs. 3(c) and 3(d), shows that the phase reconstruction using CS-COACH is more uniform in comparison to the phase evaluated by the open aperture holography, as should be the case for flat surfaces. As a result, the height of the resolution target element estimated using each method from the corresponding phase delay was found to be more accurate for CS-COACH (6.8% higher height from the nominal value of 250 nm) than open aperture holography (9.2% higher height from the nominal value of 250 nm). Results of this comparative quantitative analysis between the two approaches are summarized in Table 1.

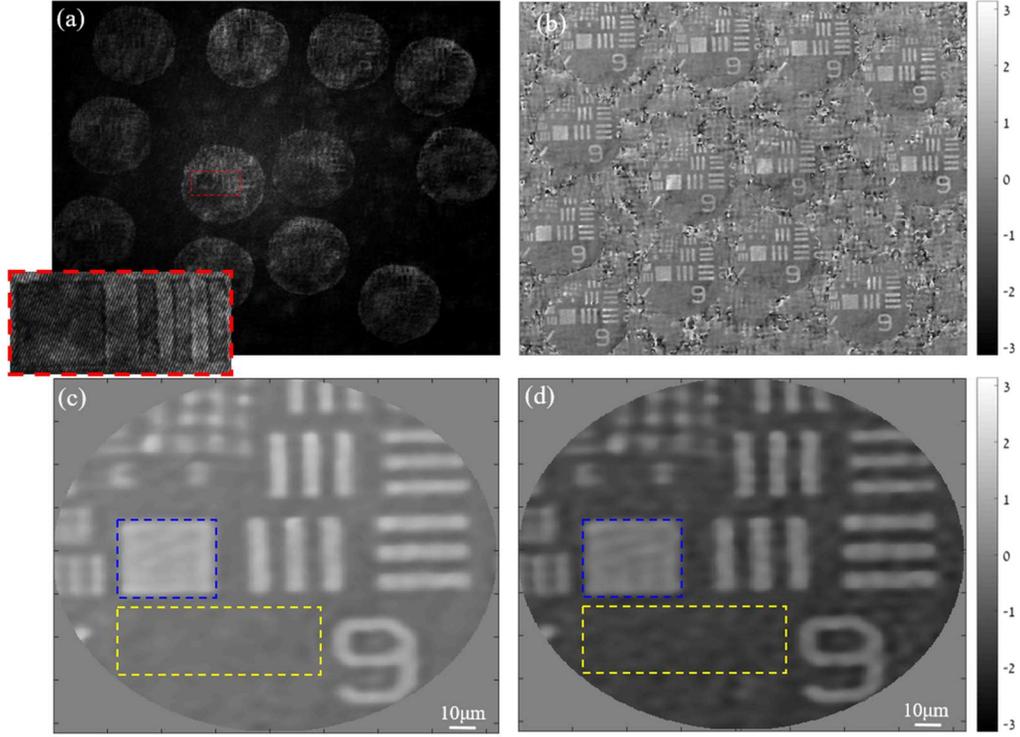

Fig. 4. Comparison between CS-COACH and standard open aperture quantitative phase imaging. (a) Off-axis hologram recorded using CS-COACH containing image replications and (b) the phase of its filtered signal after phase background subtraction. Quantitative phase imaging of 1951 USAF pure phase resolution chart using (c) CS-COACH and (d) standard open aperture holography.

Table 1. Properties of quantitative phase imaging by open aperture holography and CS-COACH

|  | Open aperture | CS-COACH |
| --- | --- | --- |
| Phase element MSE (blue dashed square) | 1.97e-3 | 1.92e-3 |
| Substrate MSE (yellow dashed square) | 10.86e-4 | 7.06e-4 |
| Phase element height [nm] | 273 | 267 |

To further examine CS-COACH capability, we image phase objects having continuous refractive index variation and higher physical extent, compared to the previous experiment. Since the measured quantity in quantitative phase microscopy is the phase values, the significance of using accurate method for objects having extremely small refractive index variations is more pronounced. As will be demonstrated next, errors in phase measurement induce higher deviations from the real values (sphere diameter for the following experiment) in the case of low refractive index variation. For this experiment, glass slide of Polystyrene microspheres (Focal Check, 6μm diameter) sealed within a mounting medium was placed at the object plane of the CS-COACH system. The refractive indices of the Polystyrene microspheres of 6 μm diameter and the mounting medium were 1.6 and 1.56, respectively, leading to a maximal phase delay of 2.38 radians across the sample. To evaluate and compare the system performance, we extracted the phase information of the microspheres sample twice, first by CS-COACH and then by standard open aperture holography, similarly to the previous experiment. Figs. 5(a) and 5(b) illustrate the reconstructed phase of a partial area from the sample containing three microspheres, recorded by regular off-axis holography and by the suggested method, respectively. The SNR of the reconstructed phase, evaluated from the phase of the sample background, was found to be slightly better in the case of the reconstruction using CS-COACH. However, better performance of CS-COACH in this case is expressed more explicitly in the comparison of the cross-sections of the three microspheres phase reconstruction, shown in Figs. 5(c)-5(f). The horizontal cross-sections of the three microsphere phase distributions reconstructed using regular off-axis holography and CS-COACH are displayed in Figs. 5(c) and 5(d), respectively. Plot colors refer to the colors circulating each microsphere in Figs. 5(a)-5(b) and the dashed line indicates the mean value of the reconstructed phase background. Figs. 5(e) and 5(f) are the vertical cross-sections of the three microsphere phase distributions reconstructed using regular off-axis holography and CS-COACH, respectively. One can observe the reconstruction quality differences between the two approaches. While the phase cross-sections of the different microspheres from the open aperture holography [Figs. 5(c) and 5(e)] exhibit low consistency and have low symmetry, the phase cross-sections emerging from the CS-COACH [Figs. 5(d) and 5(f)] have low variations among the different spheres and exhibit good symmetry, as would be expected from similar, spherical structures. Moreover, the mean value of the background phase in the open aperture cross-sections is higher than the phase delay of some of the microspheres, which is unlikely to happen in the given sample. On the contrary, this mean value in the CS-COACH is below all the microspheres phase value, setting a reasonable phase value for the immersion medium. As a final merit to compare the two methods, we quantified the three microspheres diameter based on the phase reconstructions of the Polystyrene and the immersion medium from Figs. 5(a)-5(b). The microspheres diameter is estimated by subtracting the mean phase value of the immersion from the maximal phase delay at the center of the sphere, and converting it to physical length by dividing it by the product between the wavenumber and the difference between the refractive indices of the two media. The results are summarized in Table 2 and show good agreement with the manufacturer reported diameter of 6µm in the case of CS-COACH for the three microspheres, whereas the standard method exhibits high deviations from that value.

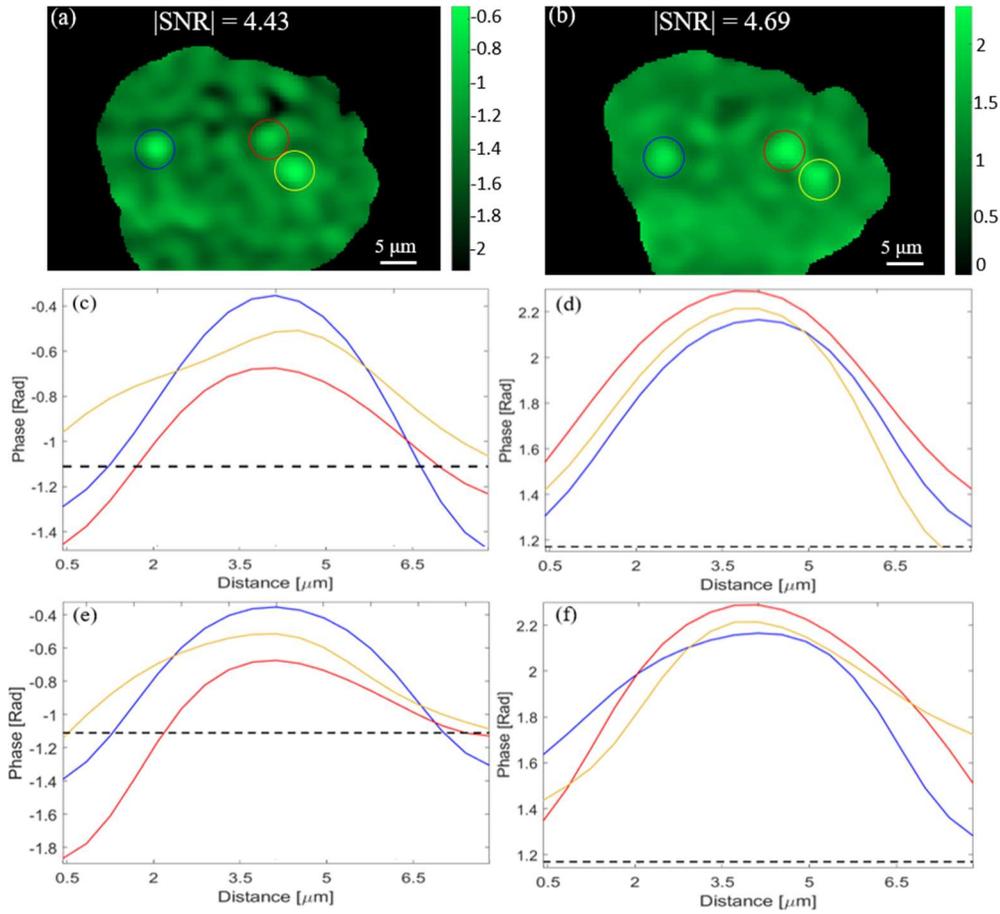

Fig. 5. Qualitative and quantitative comparison between reconstructed phase using (a) conventional off-axis holography and (b) suggested CS-COACH system. (c),(d) Vertical and (e),(f) horizontal cross-sections of the microspheres reconstructed phase using (c),(e) conventional off-axis holography and (d),(f) using CS-COACH. Plot colors are corresponding to the circles colors from (a) and (b), dashed lines stand for the mean value of the background phase.

Table 2. Microspheres diameters calculated from open aperture holography and CS-COACH

|  | Open aperture | CS-COACH |
| --- | --- | --- |
| Sphere diameter (blue circle) [μm] | 3.82 | 5.27 |
| Sphere diameter (red circle) [μm] | 2.2 | 5.63 |
| Sphere diameter (yellow circle) [μm] | 3.0 | 5.26 |

### 3.3 Extended FOV of phase objects

Combination of the two features of CS-COACH demonstrated thus far, namely, threefold increased FOV in both axes and accurate quantitative phase imaging of transparent specimens, can be of a great interest in biomedical microscopy, especially in cases where movement or scanning of the object are undesired [4,26]. Here we show that this combination can be easily implemented in the CS-COACH system, by displaying the CPM similar to the first experiment of extending the FOV on the SLM and effectively multiplex information regarding the object portion located outside the regular system FOV into the recorded interferogram. Fig. 6(a) is the acquired off-axis hologram that contains two kinds of image replications, from two different regions of a Polystyrene microspheres sample separated by a distance larger than the FOV of a regular system. The phase of the filtered hologram free of the bias and twin image terms is shown in Fig. 6(b), where the multiple phase replications of the two different regions are recognized. On that basis, we created the object hologram for the Polystyrene microspheres sample by zero-padding the filtered signal to match its size to the synthetic PSH, which is three times larger than the area size of the image sensor in both axes [Fig. 6(c)]. By cross-correlating the object hologram with the synthetic PSH we reconstruct the phase information of the specimen parts located outside the regular system FOV. Fig. 6(d) illustrates that two sample regions outside the FOV of an equivalent, open-aperture system were recovered, one contains two neighboring microspheres and the other contains a single microsphere as shown in the accompanied insets. The goal of this demonstration is to show the FOV extension capability of CS-COACH in the case of phase objects, rather than to show the accurate phase measurement capability as in the previous experiment. Therefore, microspheres of larger diameter compared to the previous experiment were chosen, in order to enhance the visibility of the reconstructed phase without conducting quantitative analysis.

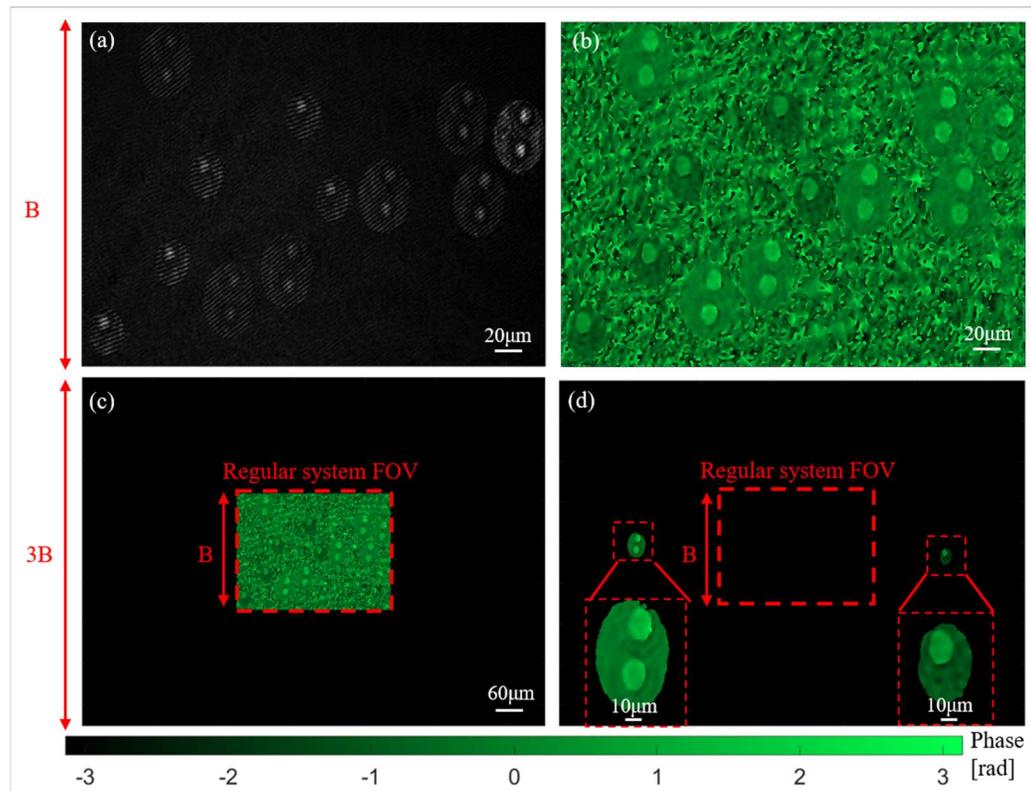

Fig. 6. Reconstruction of two different regions from Polystyrene microspheres slide which are located outside the FOV of a regular system. (a) Recorded off-axis hologram and its (b) filtered phase. (c) Phase of the zero padded object hologram and (d) the corresponding phase object reconstruction after cross correlation with the synthetic PSH.

## 4. Discussion and Conclusions

Designing a digital holographic apparatus involves a well-known tradeoff between having a high space-bandwidth product at the expanse of a low temporal bandwidth, or vice-versa [27,28]. Usually, off-axis holography has higher temporal bandwidth while phase-shifting holography systems are higher in terms of spatial bandwidth. In the presented CS-COACH method, however, we manage to perform off-axis hologram recording in a single camera shot while capturing spatial information from outside of the sensor limited size. Standard off-axis holography systems tend to maximize their temporal bandwidth by recording a hologram from a single camera shot, while their FOV is limited only to the active area of the image sensor. Since CS-COACH is capable of increasing the FOV of an equivalent, open-aperture imaging system, this capability is translated to increasing the amount of information captured by the system, while still the holograms are captured by a single camera shot. Figs. 3 and 6 demonstrates that CS-COACH is simultaneously more rapid imager than an equivalent phase-shifting holography system, and has wider FOV than an equivalent off-axis holography system. However, these benefits are achieved at the expense of some reduction in the SNR as is evident from the comparison of Figs. 3(c) and 3(d). Furthermore, in CS-COACH the FOV can be extended beyond the active area of the camera, but the non-overlap condition restricts that the size of each observed object is limited by $B/(M_T\sqrt{K})$, where $B$ is the camera width, and $K$ is the number of dots of the PSH.

One of the important figures of merit in assessing the performance of an imaging system is the SNR of the reconstructed signal. As was mentioned, the present study is not the first time coded apertures are used to increase the FOV of an imaging system [12]. The other attempts, however, exhibited a relatively poor SNR and the emerging temporal resolution of the methods was compromised to achieve better SNR. Here we present decent SNR performance based on a single camera exposure, maximizing the time-bandwidth product of the method. We believe the improved SNR of CS-COACH in comparison to other coded aperture imaging systems is mainly attributed to two mechanisms. First, cross-correlating the object hologram containing image replications with the PSH, containing an array of randomly distributed dots, can be considered as a kind of averaging process, in which each image replication is a single observation and the number of random dots is the number of observations in the ensemble. This averaging process eventually increase the signal strength over the background noise. The second beneficial mechanism of the current approach arises from the random distribution of the dots in the PSH of the optical system, which increases the complexity during the cross-correlation and therefore suppress the background noise as much as this complexity is increased [19]. Naturally, suppression of the background noise is of a high impact on the resulting SNR of the reconstructed image. Fig. 3 is a good evidence of the enhanced SNR performance of CS-COACH in imaging an amplitude object, however this characteristic has another major implication in imaging phase objects. Figs. 4 and 5 and Tables 1 and 2 emphasize the CS-COACH capability to accurately measure the quantitative phase of the examined object in comparison to an equivalent, well established technique of off-axis holography with open aperture. With CS-COACH we were able to quantify the diameter of the Polystyrene microspheres from their emerging phase profiles with a mean error of 10%, while the corresponding open aperture holography error was 5 times higher. This is in addition to the consistency of the microspheres phase profiles and higher SNR in the case of CS-COACH, which are presented in Fig. 5.

All the above-mentioned advantages lead us to conclude that CS-COACH can be regarded as a practical approach to quantitative phase microscopy. It is important to note the non-trivial imaging conditions exist in this experiment. Note that since the variations in the OT of the examined object are lower than the illumination wavelength, the system is more prone to phase noise. Therefore, an accurate extraction of the object's quantitative phase is a serious challenge in comparison to other cases where the changes in the OT of the examined object are larger than the illumination wavelength, simply because the dynamic range is larger.

Another beneficial aspect of CS-COACH is the redundancy of the pre-imaging calibration process. Coded aperture imaging systems usually involve a process in which the reconstructing function is experimentally acquired in a one-time guidestar calibration for every system configuration. In this study, we have demonstrated that the reconstructing function can be rendered in the computer based on the knowledge of the system magnification and of the distribution of the dot array. Although the one-time calibration process has no influence on the imaging durations of the system, avoiding the calibration can be beneficial for two main reasons. First, it provides another level of modularity to the system in cases where one or more optical elements must be changed, without the need for a new calibration process. Second, in the present case of the extended FOV, the calibration process is quite long consisting of placing a point object at 9 different locations of the object plane, experimentally record and computationally combine all the 9 sub-PSHs into a single synthetic PSH [12]. The alternative of digitally render the corresponding PSH in the computer releases the experimenter from this one-time, however, long process.

As is shown in this study, CS-COACH provides a more reliable quantification of the object's phase in comparison to an equivalent open-aperture system. This advantage is more important in cases where the imaged phase is more prone to noise due to the object's thin OT of less than a wavelength – a very common situation in metrology and biomedical imaging.

**Funding**

European Commission (777222) for ATTRACT project; Israel Science Foundation (1669/16); Israel Ministry of Science and Technology.

**Disclosures**

The authors declare no conflicts of interest.